\documentclass[universe,article,accept,moreauthors,10pt,a4paper]{mdpi}
\firstpage{1} 
\articlenumber{x}
\doinum{10.3390/------}
\pubvolume{xx}
\pubyear{2018}
\copyrightyear{2018}
\externaleditor{Academic Editor: name}
\history{Received: date; Accepted: date; Published: date}

\Title{Phenomenological Equation of State of Strongly Interacting Matter
  with First-Order Phase Transitions and Critical Points}

\Author{Stefan Typel $^{1,2}$* 
and David Blaschke $^{3,4,5}$}

\AuthorNames{Stefan Typel and David Blaschke}

\address{%
$^{1}$ \quad Institut f\"{u}r Kernphysik, Technische Universit\"{a}t Darmstadt,
Schlossgartenstra\ss{}e 9, 64289 Darmstadt, Germany; stypel@ikp.tu-darmstadt.de\\
$^{2}$ \quad GSI Helmholtzzentrum f\"{u}r Schwerionenforschung GmbH, Planckstra\ss{}e 1,
64291 Darmstadt, Germany; s.typel@gsi.de\\
$^{3}$ \quad Institute of Theoretical Physics, University of Wroclaw, Pl. M. Borna 9,
50-204 Wroclaw, Poland; david.blaschke@gmail.com\\
$^{4}$ \quad Bogoliubov Laboratory for Theoretical Physics, Joint Institute of Nuclear 
Research, 141980 Dubna, Russia\\
$^{5}$ \quad National Research Nuclear University (MEPhi), Kashirskoe shosse 31, 
115409 Moscow, Russia
}

\corres{Correspondence: s.typel@gsi.de; Tel.: +49-6159-71-2744}

\abstract{An extension of the relativistic density functional approach to the equation of~state for
  strongly interacting matter is suggested that generalizes a recently developed
  modified excluded-volume mechanism
  to the case of temperature- and density-dependent available-volume fractions.
  A~parametrization of this dependence is presented for which, at low temperatures and
  suprasaturation densities, a first-order phase transition is obtained. It changes for increasing
  temperatures to a crossover transition via a critical endpoint.
  This provides a benchmark case for studies of the role of such a point in hydrodynamic
  simulations of ultrarelativistic heavy-ion collisions.
  The approach is thermodynamically consistent and extendable to finite isospin asymmetries
  that are relevant for simulations of neutron stars, their mergers, and core-collapse supernova explosions.
}


\begin{document}


\section{Introduction}

{The simulation of astrophysical phenomena, such as core-collapse supernovae (CCSN)
or neutron-star (NS)} mergers,
requires a careful modeling of strongly interacting matter in a wide range of 
densities and temperatures. The same applies to the theoretical description of heavy-ion
collisions (HIC) that study compressed baryonic matter in the laboratory
from low to high beam energies. The properties of such matter are represented
by the equation of state (EoS) that provides information on pressure, entropy,
energies, and other thermodynamic variables of interest.

A particular feature of QCD matter
is the supposed phase transition (PT) from hadronic matter to quark matter
when density or temperature increase to sufficiently high values.
A strong first-order PT
could allow for the existence of a third branch of compact stars and the
occurrence of the twin-star phenomenon \cite{Alford:2013aca,Benic:2014jia,Alvarez-Castillo:2017qki}.
Signals of the PT might also have direct consequences in dynamical
processes when matter in the quark phase expands and cools down, e.g., the release of
a second neutrino burst in CCSN \cite{Sagert:2008ka,Fischer:2008rh,Fischer:2010wp}.
For a recent review on the role of the EoS in CCSN simulations, see \cite{Fischer:2017zcr}. 

The theoretical description of the hadron--quark PT in strongly interacting
matter often relies on a~construction employing different models for the two phases.
With such an approach, the coexistence line of the first-order PT will usually
connect a point on the zero-temperature axis at finite baryon chemical potential $\mu_{B}$
with a point at finite temperature $T$ on the zero baryon chemical potential axis. 
By this construction, the QCD hadron--quark PT is of first order in the whole temperature--density plane,
see for instance \cite{Blaschke:2010ka}.
However, from lattice QCD studies, it is known that there is a~smooth crossover at $\mu_{B}=0$
with increasing $T$ \cite{Bazavov:2014pvz,Borsanyi:2013bia}, so at least one critical point at finite $\mu_{B}$ and $T$ is expected to exist.
Other possibilities are that the character of the transition is crossover all over the QCD phase diagram
\cite{Bratovic:2012qs}
or, as is advocated in studies of the BEC-BCS crossover transition in low-temperature QCD, 
that a second critical endpoint exists \cite{Baym:2017whm,Abuki:2010jq,Hatsuda:2006ps}.
Since lattice QCD studies are presently incapable of exploring the EoS
close to the presumed critical point with much confidence, unified models are needed 
that can account for the existence of these features, see, e.g., \cite{Klahn:2015mfa}.
There are dedicated microscopic models available that
incorporate the major expected features
in the QCD phase diagram, e.g., chiral mean-field models~\cite{Dexheimer:2009hi}
or parity-doublet quark--hadron models~\cite{Mukherjee:2016nhb}.
Simulations of CCSN or HIC that are based on a hydrodynamic description
of matter during dynamical evolution use the thermodynamic properties of
matter encoded in the EoS as an input. Such~data can be provided 
by phenomenological models that
do not need to incorporate all of the details of the underlying physics.

In this work, a novel approach is introduced to provide a phenomenological EoS of
baryonic matter that exhibits a first-order PT and a critical point at
densities and temperatures expected in QCD matter. The parameters of the model can be
adjusted to place the coexistence line at arbitrary positions in the phase diagram.
The description uses an extension of a relativistic energy density functional for
hadronic matter assuming a medium-dependent change in the number of degrees of freedom.
This approach employs a recently developed version of a modified 
excluded-volume (EV) mechanism
that gives a thermodynamically consistent EoS with nuclear matter properties that are
consistent with present constraints. Here, we concentrate on the hadron--quark transition
but not on the liquid--gas PT, which is also contained in our model.
The model allows us to study the PT for arbitrary isospin asymmetries; however,
only isospin-symmetric matter is considered in this first exploratory study for simplicity.
In the present work, no attempt was made to reproduce the EoS of QCD matter at vanishing
baryon chemical potential obtained in lattice QCD studies.
With appropriately chosen EV parameters,
the crossover transition with increasing temperature can be well modeled,
even for imaginary chemical potentials,
e.g., in a hadron resonance gas model \cite{Vovchenko:2017xad}.
With improved parametrizations, the structure of the phase diagram in the full space of
variables, i.e., temperature, baryon density/chemical potential, and isospin asymmetry,
can be investigated in the future.

The theoretical formalism of the model is presented in Section \ref{sec:theo}, which includes the main equations that define the relevant thermodynamic quantities in
Section \ref{sec:redf}. In Section \ref{sec:para}, details of the parametrization
of the interaction and of the effective degeneracy factors are given. They account for the change
in the number of degrees of freedom with density and temperature.
The phase transitions are explored in Section \ref{sec:res} for isospin-symmetric matter.
Conclusions follow in Section \ref{sec:con}.

\section{Theoretical Model}
\label{sec:theo}

The theoretical description of strongly interacting matter in the present work is
adapted from the model introduced in \cite{Typel:2016srf}. It combines
a relativistic mean-field (RMF) approach for hadronic matter 
with density-dependent nucleon--meson
couplings and a modified EV mechanism. Here, it is sufficient to provide only
the main equations without a detailed derivation. The essential quantities that determine
the position of the PT and the critical point in the phase diagram are the
effective degeneracy factors that depend on the number densities of the particles and
the temperature. 

\subsection{Relativistic Energy Density Functional with Modified 
Excluded-Volume Mechanism}
\label{sec:redf}

The present model assumes neutrons and protons as well as their antiparticles as the basic degrees
of freedom. These particles interact by the exchange of mesons, and the model effectively describes the short-range repulsion
($\omega$ meson), the intermediate-range attraction ($\sigma$ meson), and the isospin dependence
of the nuclear interaction ($\rho$ and $\delta$ mesons), as is common of RMF models. The contribution of leptons or other degrees of freedom like nuclei, hyperons
or photons, as required for multi-purpose EoS for astrophysical applications
\cite{Oertel:2016bki}, is not considered here.

The nucleons $i=n,p,\bar{n},\bar{p}$ with rest masses $m_{i}$ are treated as quasi-particles
of energy,
\begin{equation}
  \label{eq:E}
  E_{i}(k) = \sqrt{k^{2}+\left(m_{i}-S_{i}\right)^{2}}+V_{i}
\end{equation}
which depends on the particle momentum $k$ and the scalar ($S_{i}$) and vector ($V_{i}$) potentials.
Denoting the particle chemical potentials with $\mu_{i}$, the contribution of the quasi-particles
to the total pressure
\begin{equation}
  \label{eq:p_tot}
  p = \sum_{i} p_{i} + p_{\rm meson} - p^{(r)}
\end{equation}
of the system can be written as
\begin{equation}
 p_{i} = T g_{i}^{\rm (eff)} \int \frac{d^{3}k}{(2\pi)^{3}} \:
 \ln \left[ 1 + \exp \left( - \frac{E_{i}(k)-\mu_{i}}{T} \right)\right]
\end{equation}
where the medium-dependent effective degeneracy factors
\begin{equation}
  \label{eq:g_eff}
  g_{i}^{\rm (eff)} = g_{i} \Phi_{i}
\end{equation}
are a product of the usual degeneracy factor $g_{i}=2$ for nucleons and the
available-volume fraction $\Phi_{i}$, which is defined in Section \ref{sec:para}.

The meson contribution 
\begin{equation}
  \label{eq:p_meson}
 p_{\rm meson} = \frac{1}{2} \left( C_{\omega} n_{\omega}^{2} + C_{\rho} n_{\rho}^{2} 
 - C_{\sigma} n_{\sigma}^{2}  - C_{\delta} n_{\delta}^{2}\right)
\end{equation}
to the total pressure in Equation (\ref{eq:p_tot})
contains the coupling factors of the mesons
\begin{equation}
  \label{eq:C}
  C_{j} = \frac{\Gamma_{j}^{2}}{m_{j}^{2}}
\end{equation}
{given as a ratio of the density-dependent coupling functions $\Gamma_{j}$ and the
meson masses $m_{j}$. The source densities}
\begin{equation}
  \label{eq:source_v}
  n_{j} = \sum_{i} g_{ij} n_{i}^{(v)}
\end{equation}
for vector mesons ($j=\omega,\rho$) and
\begin{equation}
  \label{eq:source_s}
  n_{j} = \sum_{i} g_{ij} n_{i}^{(s)}
\end{equation}
for scalar mesons ($j=\sigma,\delta$)
in Equation (\ref{eq:p_meson}) are obtained
from the quasi-particle vector densities
\begin{equation}
  \label{eq:n_v}
  n_{i}^{(v)} = g_{i}^{\rm (eff)}  \int \frac{d^{3}k}{(2\pi)^{3}} \: f_{i}(k)
\end{equation}
and scalar densities
\begin{equation}
  \label{eq:n_s}
  n_{i}^{(s)} = g_{i}^{\rm (eff)}  \int \frac{d^{3}k}{(2\pi)^{3}} \: f_{i}(k)
 \frac{m_{i}-S_{i}}{\sqrt{k^{2} + \left( m_{i} - S_{i}\right)^{2}}}
\end{equation}
with the Fermi-Dirac distribution function
\begin{equation}
 f_{i}(k) = \left[ \exp \left( \frac{E_{i}(k)-\mu_{i}}{T} \right) + 1 \right]^{-1} \: .
\end{equation}

The scaling factors
\begin{eqnarray}
  g_{n\omega}=g_{p\omega}=-g_{\bar{n}\omega}=-g_{\bar{p}\omega} & = & 1 \\
  g_{n\rho}=-g_{p\rho}=-g_{\bar {n}\rho}=g_{\bar{p}\rho} & = & 1 \\
  g_{n\sigma}=g_{p\sigma}=g_{\bar{n}\sigma}=g_{\bar{p}\sigma} & = & 1 \\
  g_{n\delta}=-g_{p\delta}=g_{\bar{n}\delta}=-g_{\bar{p}\delta} & = & 1 
\end{eqnarray}
in Equations (\ref{eq:source_v}) and (\ref{eq:source_s}) determine the
coupling between mesons and nucleons. They also appear
in the vector potential
\begin{equation}
  \label{eq:V}
  V_{i} = C_{\omega} g_{i\omega} n_{\omega} + C_{\rho} g_{i\rho} n_{\rho} 
  + B_{i} V_{\rm meson}^{(r)} + V_{i}^{(r)}
\end{equation}
and the scalar potential
\begin{equation}
  \label{eq:S}
 S_{i} = C_{\sigma} g_{i\sigma} n_{\sigma} + C_{\delta} g_{i\delta} n_{\delta} + S_{i}^{(r)}
\end{equation}
in the quasi-particle energy (Equation (\ref{eq:E})). The rearrangement potential
\begin{equation}
 V_{\rm meson}^{(r)} = \frac{1}{2} \left( 
   C_{\omega}^{\prime} n_{\omega}^{2}  + C_{\rho}^{\prime} n_{\rho}^{2} 
 - C_{\sigma}^{\prime} n_{\sigma}^{2}  - C_{\delta}^{\prime} n_{\delta}^{2}\right)
\end{equation}
contributes to the vector potential (Equation (\ref{eq:V})) because the couplings $\Gamma_{j}$
in Equation (\ref{eq:C}) are assumed to depend on the baryon density
$ n_{B} = \sum_{i} B_{i} n_{i}^{(v)} $,
where $B_{i}=g_{i\omega}$ is the baryon number of particle $i$, and the quantities
$C_{j}^{\prime} = dC_{j}/d n_{B}$
are the derivatives of the coupling factors.

The dependence of the available-volume fractions $\Phi_{i}$ in the effective
degeneracy factor (Equation~(\ref{eq:g_eff})) on the vector or scalar
quasi-particle densities (\ref{eq:n_v}) and (\ref{eq:n_s}) also generates
rearrangement contributions
\begin{equation}
  V_{i}^{(r)}  =  - \sum_{j} p_{j} \frac{\partial \ln \Phi_{j}}{\partial n_{i}^{(v)}}
\end{equation}
and
\begin{equation}
  S_{i}^{(r)}  =   \sum_{j} p_{j} \frac{\partial \ln \Phi_{j}}{\partial n_{i}^{(s)}}
\end{equation}
in the potentials (Equations (\ref{eq:V}) and (\ref{eq:S})), respectively. Furthermore,
there is a rearrangement term
\begin{equation}
 p^{(r)} = p^{(r)}_{\rm meson} + p^{(r)}_{\Phi}
\end{equation}
in the total pressure (Equation (\ref{eq:p_tot})) with two contributions from
the density dependence of the couplings
\begin{equation}
  p^{(r)}_{\rm meson} 
  = -V^{(r)}_{\rm meson} n_{B}
\end{equation}
and 
\begin{equation}
  p^{(r)}_{\Phi} = \sum_{i} \left( n_{i}^{(s)} S_{i}^{(r)}
  - n_{i}^{(v)} V_{i}^{(r)}\right) 
\end{equation}
from the EV effects.

The free energy density of the system
\begin{equation}
 f  = \sum_{i} \mu_{i} n_{i}^{(v)} - p
\end{equation}
is obtained with the total pressure (Equation (\ref{eq:p_tot})) and the chemical potentials of the
particles $\mu_{i}$. They are not independent since, for nucleons with baryon number $B_{i}$
and charge number $Q_{i}$, they are given~by
\begin{equation}
   \mu_{i} = B_{i} \mu_{B} + Q_{i} \mu_{Q}
\end{equation}
with the baryon chemical potential $\mu_{B}$ and the charge chemical potential $\mu_{Q}$.
Only the latter two are independent quantities.
For the internal energy density
\begin{equation}
  \varepsilon = f+Ts
\end{equation}
the entropy density
\begin{equation}
  s = - \sum_{i} g_{i}^{\rm (eff)}  \int \frac{d^{3}k}{(2\pi)^{3}} \: \left[ 
 f_{i} \ln f_{i} + \left(1-f_{i}\right) \ln \left(1-f_{i}\right) \right]
 + \sum_{i} p_{i} \frac{\partial \ln \Phi_{i}}{\partial T}
\end{equation}
is needed. Besides the standard contribution depending on the distribution functions
$f_{i}$, there is a term from the possible temperature dependence of the
available-volume fractions $\Phi_{i}$.
In order to guarantee the third law of thermodynamics, i.e.,\
$\lim_{T \to 0} s = 0$, the temperature derivative of the available-volume fractions has to vanish
for $T \to 0$. After solving the equations above for a given
$T$, $\mu_{B}$, and $\mu_{Q}$, a fully consistent thermodynamic EoS is obtained. For practical
purposes, however, the baryon density $n_{B}$ and the hadronic charge fraction
\begin{equation}
  Y_{q} = \frac{\sum_{i} Q_{i} n_{i}^{(v)}}{n_{B}}
\end{equation}
are used as independent variables instead of $\mu_{B}$ and $\mu_{Q}$.

A possible shortcoming of models that consider EV effects is the potential
appearance of a~superluminal speed of sound in certain regions of the space
of thermodynamic variables, see, e.g., \cite{Rischke:1991ke,Satarov:2009zx}.
This causality constraint has to be checked
case by case depending on the specific implementation of the EV mechanism.

\subsection{Available-Volume Fractions and Model Parameters}
\label{sec:para}

For a quantitative evaluation of the EoS in the present approach, the functional forms
of the meson--nucleon couplings $\Gamma_{j}$ and the available-volume fractions $\Phi_{i}$
have to be specified as well as all parameters. Here, we use the masses of nucleons and mesons
and the coupling functions of the DD2 parametrization presented in 
\cite{Typel:2009sy}. It only considers $\omega$, $\rho$, and $\sigma$ mesons 
for the effective description of the nuclear interaction but not the $\delta$ mesons.
The parameters were obtained by fitting observables (binding energies, radii, etc.) of selected nuclei.
With this set, the EoS of nuclear matter at zero temperature exhibits
characteristic nuclear matter parameters, e.g.,
the saturation density ($n_{\rm sat}=0.149065$~fm$^{-3}$),
binding energy at saturation ($B=16.02$~MeV),
incompressibility ($K=242.7$~MeV),
symmetry energy ($J=32.73$~MeV), and slope ($L=57.94$~MeV), that are
consistent with modern constraints from experiment and theory.

EV effects are frequently employed to describe an effective repulsive interaction
between particles, in particular in calculations of the EoS in the framework of
hadron resonance gas models, see, e.g., \cite{Vovchenko:2016rkn}.
For zero baryon density, a comparison of the
resulting EoS with results from lattice QCD calculations can be used to fix the 
volume parameters.
If a finite volume $v_{i}$ is attributed to
each particle $i$, the available volume for the motion of the particle 
is reduced from the total
system volume $V$ to $V\Phi^{(cl)}$ with the classical available-volume fraction
\begin{equation}
  \Phi^{(cl)} = 1-\sum_{i}v_{i}n_{i}^{(v)} \: .
\end{equation}

Clearly, there is a limiting density above which a compression of the system becomes
impossible. In general, the volumes and available-volume fractions can depend on the
particle species, and the EV mechanism can be used to suppress
particles, e.g., nuclei, in a mixture, causing them to disappear above a~certain density, see, e.g., \cite{Hempel:2009mc} for applications to the low-temperature and low-density
EoS in astrophysical simulations.

In \cite{Typel:2016srf}, the interpretation of the EV mechanism was changed
by moving the factor $\Phi_{i}$ from the system volume $V$ to the degeneracy 
factor $g_{i}$ as in Equation (\ref{eq:g_eff}) and allowing the available-volume 
fractions to be arbitrary functions of the particle densities and temperature. The~medium dependence of the effective
degeneracy factors is interpreted as a change in the effective number of degrees
of freedom. A decrease in $g_{i}^{\rm (eff)}$ has the effect of a repulsive interaction
between the particles, whereas an increase can be seen as the action of an 
attractive interaction, c.f., the softening of the nuclear EoS when hyperons are
included, see, e.g., \cite{Chatterjee:2015pua,SchaffnerBielich:2008kb}.
This freedom leaves the room to modify the properties of an EoS in a favored way.

In the present application of the modified EV mechanism, the available-volume
fraction is defined to be identical for all particles $i$ as
\begin{equation}
  \label{eq:Phi}
 \Phi_{i} = 1 + s g_{1}(T) \theta(x) \exp \left( -\frac{1}{2x^{2}}\right) 
\end{equation}
depending on the temperature $T$ and an auxiliary quantity 
\begin{equation}
  \label{eq:x}
 x = v \left( \sum_{j} B_{j} n_{j}^{(v)} - g_{8}(T) n_{\rm cut}\right)
\end{equation}
depending on $T$ and the quasi-particle vector densities $n_{i}^{(v)}$
with parameters $s$, $v$, and a cutoff density of $n_{\rm cut}$. The $\theta$ function
in Equation (\ref{eq:Phi}) guarantees that $\Phi_{i} = 1$ for $x\leq 0$ and the EV
mechanism has no effect on the EoS. The functions $g_{1}$ in Equation (\ref{eq:Phi})
and $g_{8}$ in Equation (\ref{eq:x}) are defined as
\begin{equation}
  \label{eq:g_t}
 g_{t}(T) = \theta\left( T_{0} - T\right) 
 \exp \left[ - \frac{t}{2} \left( \frac{T}{T_{0}-T}\right)^{2} \right]
\end{equation}
with parameters $t$ and $T_{0}$. In the limit $T\to 0$, the function $g_{t}$ approaches one
and it decreases with increasing temperature.
Furthermore, the derivatives $\partial \Phi_{i}/\partial T$ approach zero for $T \to 0$,
as required for the thermodynamic consistency, because of the choice of the function $g_{t}(T)$.
For $T \to T_{0}$, the function $g_{t}$ vanishes very smoothly
and there are no effects at higher temperatures
because $\Phi_{i}=1$.
In~order to reproduce the correct high-temperature limit, given by a Stefan--Boltzmann-type behavior, a~modification of the available-volume fractions for temperatures
well above $T_{0}$ is required. This~is left to future extensions of the model.
According to Equation (\ref{eq:x}),
the quantity $x$ is only positive for baryon densities larger than $g_{8}n_{\rm cut}$, 
a quantity that decreases with increasing temperature.
There are no artificial singularities due to the presence of the $\theta$
functions in Equations (\ref{eq:Phi}) and (\ref{eq:g_t}) because all derivatives
of the exponential functions
are zero when the arguments of the $\theta$ functions vanish.
The~actual values of the parameters for the modified EV mechanism used in the present 
study are $s=3$, $v=2$~fm$^{3}$, $T_{0} = 270$~MeV, and $n_{\rm cut} = n_{\rm sat}$ 
of the DD2 parametrization.


\section{Results}
\label{sec:res}

\begin{figure}[H]
  \centering
  \includegraphics[width=7cm]{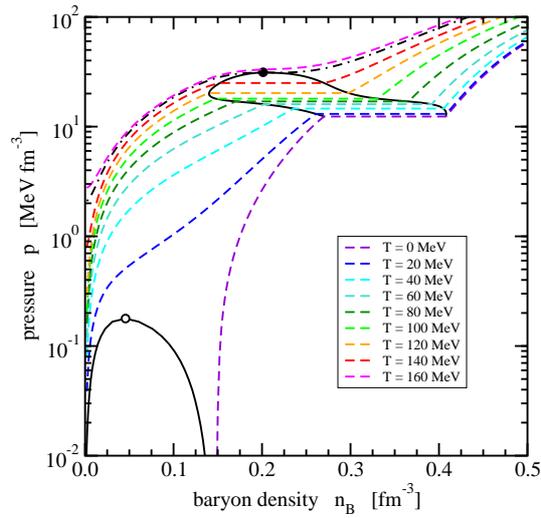}\vspace{-6pt}
  \caption{\label{fig:1}
    Isotherms in isospin-symmetric strongly interacting matter in the pressure--baryon
    density diagram at temperatures from $0$ to $160$~MeV in steps of $20$~MeV
    (dashed colored lines) and at the critical temperature $T_{\rm crit}$ of the pseudo
    hadron--quark phase transition (black dot-dashed line).
    The binodals and critical points are denoted by full black lines and a full (open) circle
    of the pseudo hadron--quark (liquid-gas) phase transition,
    respectively.}
\end{figure}

In order to illustrate the characteristic effects of the modified EV mechanism on the EoS,
we limit the presentation to the case of symmetric matter, i.e., $Y_{q}=0.5$. The pressure $p$
and baryon chemical potential $\mu_{B}$ are calculated as a function of the baryon density
$n_{B}$. Due to the increase in the available-volume fractions $\Phi_{i}$ or the
effective degeneracy factors $g_{i}^{\rm (eff)}$, a considerable
softening of the EoS is observed in a certain range of densities.
Below the critical temperature $T_{\rm crit}$, the pressure
is not a monotonous function of the baryon chemical potential for an isotherm. A Maxwell
construction is used to determine the two densities of the coexisting phases where $p$ and
$\mu_{B}$ are identical. The~pressure $p$ as a function of the baryon density $n_{B}$
is depicted in Figure \ref{fig:1} for selected temperatures.
Note that an integral part of the underlying RMF model is a detailed description of the
liquid--gas PT in nuclear matter and the formation and dissociation of nuclear clusters
in compact-star matter. For details, see, e.g., \cite{Typel:2013zna}.

In the coexistence region of the pseudo hadron--quark PT
between the low and high density phases, the pressure is constant as typical for a first-order
PT. The area of coexistence is enclosed by the binodal, and, above the critical
temperature $T_{\rm crit}\approx 155.5$~MeV, there is no PT anymore.
The peculiar shape of the binodal is a result of the specific form (\ref{eq:Phi})
of the available-volume fractions. It can be adjusted with appropriate changes in the
functional form and parameters.

\begin{figure}[H]
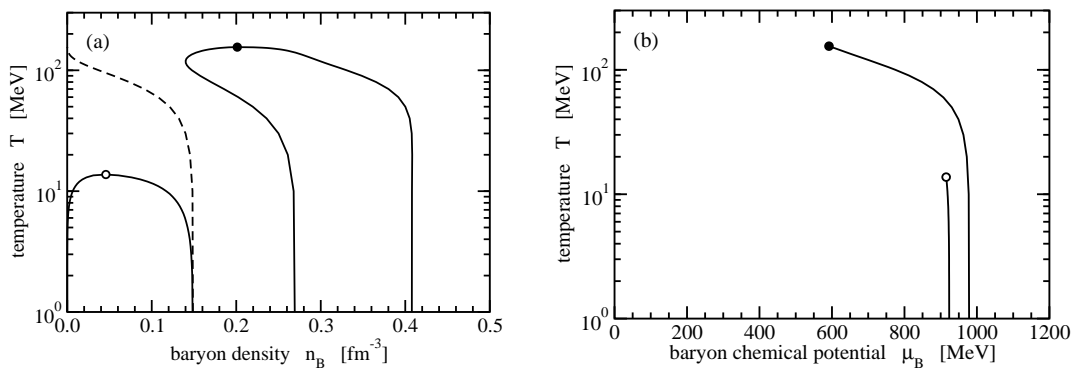

  \centering
  \includegraphics[width=6.5cm]{hq_T_nb_cut_log}
  \hspace{0.5cm}
  \includegraphics[width=6.8cm]{hq_T_mub_log}\vspace{-6pt}
  \caption{\label{fig:2}
    Binodals (full lines) and critical points (full and open circles) of isospin-symmetric strongly
    interacting matter in (\textbf{a}) the temperature--baryon density diagram and
    (\textbf{b}) the temperature--baryon chemical potential diagram. The dashed line in panel
    (\textbf{a}) separates the region without effects of the modified excluded-volume
    mechanism (lower left) from the region with effects (upper right).
    Results for the liquid--gas phase transition are shown at subsaturation densities.}
\end{figure}

The binodals of the liquid--gas and pseudo hadron--quark PT
in the temperature--baryon density plane are shown in panel (a) of 
Figure \ref{fig:2}.
At vanishing temperatures, the coexistence region of the pseudo hadron--quark PT
covers a density range from $0.270$ to $0.408$~fm$^{-3}$, 
well above the nuclear saturation density $n_{\rm sat}$. At higher temperatures, 
it moves to lower
densities with an almost constant extension in baryon density except for 
temperatures close to $T_{\rm crit}$. Here, the critical density is found as $0.201$~fm$^{-3}$, 
still above $n_{\rm sat}$.
The dashed line in panel (a) marks the boundary between regions without (lower left) and
with (upper right) effects of the modified EV mechanism in the present parametrization.
It corresponds to the condition $x=0$. There is another region in the phase diagram without 
modified EV effects at temperatures above $T_{0}$, outside the figure.

Panel (b) of Figure \ref{fig:2} depicts the lines of the first-order PT
in the temperature--baryon chemical potential diagram ending in critical points. With
increasing temperature, the baryon chemical potential at the pseudo hadron--quark
PT reduces from $979.1$ to $591.8$~MeV at the critical point. By crossing the 
transition line, an abrupt change in the density occurs that becomes continuous at the 
critical point.

\section{Conclusions}
\label{sec:con}

The extension of the modified EV approach to a density- and temperature-dependent parametrization
of the available-volume fractions
as introduced in this work was successful in achieving the main goal of this study:
As a generic structure of the QCD phase diagram, a first-order pseudo hadron--quark
phase transition at low temperatures
and a crossover for low baryon densities could be modeled that also includes a critical endpoint at 
$T_{\rm crit}= 155.5$ MeV and $\mu_{B,{\rm crit}}=591.8$ MeV.
Other patterns of the QCD phase diagram that have been theoretically motivated could also be modeled
within the present approach. 
Further extensions of the model are straightforward. They include the extension to a larger number
of components of the hadron resonance gas in the underlying RMF model and an isospin dependence of the EV model.
It would be worthwhile to study further thermodynamic quantities
such as the speed of sound, heat capacities, or the susceptibilities in such an enlarged model.
It would also be interesting to elaborate on a parametrization that would result in a second endpoint at
low temperatures, as suggested by Hatsuda et al.\ \cite{Hatsuda:2006ps}.
The so-generalized parametrization of the QCD EoS can be used in Bayesian analysis studies
for astrophysical applications pertaining to compact stars \cite{Steiner:2010fz,Alvarez-Castillo:2016oln},
their mergers, and core-collapse supernova explosions, as well as heavy-ion collisions analogous
to those studied in \cite{Pratt:2015zsa}.

\vspace{6pt} 


\acknowledgments{This work was supported by the Russian Science Foundation under contract number 17-12-01427.
S.T. was supported in part by the DFG through grant No.\ SFB1245.}

\authorcontributions{D.B. and S.T. conceived and designed the concept of this work.
  S.T. worked out the theoretical framework, performed the numerical calculations and
  wrote the main body of the paper. 
  D.B. contributed abstract, discussion, conclusions and references.}

\conflictsofinterest{The authors declare no conflict of interest. 
} 

\abbreviations{The following abbreviations are used in this manuscript:\\

\noindent 
\begin{tabular}{@{}ll}
CCSN & core-collapse supernova\\
EoS & equation of state\\
EV & excluded-volume\\
HIC & heavy-ion collision\\
NS & neutron star\\
PT & phase transition\\
QCD & quantum chromodynamics\\
RMF & relativistic mean-field
\end{tabular}}

\appendixtitles{no} 
\appendixsections{multiple} 


\reftitle{References}


\externalbibliography{yes}
\bibliography{eos_mev}


\end{document}